\documentclass[aps,prl,amsmath,amssymb,superscriptaddress,floatfix,showpacs,twocolumn]{revtex4-1}

\usepackage{graphicx}
\usepackage{epsfig}
\usepackage{epstopdf}
\usepackage{subfigure}
\usepackage{tikz}

\usepackage{dcolumn}
\usepackage{bm}
\usepackage{latexsym}
\usepackage{amssymb}
\usepackage{amsfonts}
\usepackage{amsmath}
\usepackage{color}

\usepackage[colorlinks,bookmarks=false,citecolor=blue,linkcolor=red,urlcolor=blue]{hyperref}
\usepackage{verbatim}
\usepackage{wasysym}
\usepackage{booktabs}

\usepackage{ulem} 

\def\be{\begin{equation}}
\def\ee{\end{equation}}

\def\vec{\mathbf}
\def\mc{\mathcal}

\usepackage{url}

\usepackage{color}

\definecolor{darkblue}{rgb}{0,0.02,0.45}
\definecolor{darkred}{rgb}{0.45,0.02,0}

\def\cuse{Cu$_2$OSeO$_3$}

\newcommand{\bea}{\begin{eqnarray}}
\newcommand{\eea}{\end{eqnarray}}
\newcolumntype{.}{D{.}{.}{-1}}

\def\vec{\mathbf}
\def\mc{\mathcal}

\allowdisplaybreaks

\usepackage[colorlinks,bookmarks=false,citecolor=blue,linkcolor=red,urlcolor=blue]{hyperref}

\usepackage{times}

\begin{document}

\title{Establishing the fundamental magnetic interactions in the chiral skyrmionic Mott insulator Cu$_2$OSeO$_3$ by terahertz electron spin resonance}

\author{M.~Ozerov}\email{m.ozerov@hzdr.de}
\affiliation{Dresden High Magnetic Field Laboratory (HLD), Helmholtz-Zentrum Dresden-Rossendorf, D-01328, Germany}

\author{J.~Romh\'{a}nyi}
\author{M.~Belesi}
\affiliation{Leibniz Institute for Solid State and Materials Research, IFW Dresden, D-01069, Germany}

\author{H.~Berger}
\author{J.-Ph.~Ansermet}
\affiliation{Institut de Physique de la Mati\'{e}re Condens\'{e}e, Ecole Polytechnique F\'{e}d\'{e}rale de Lausanne, Station 3, CH-1015 Lausanne-EPFL, Switzerland}

\author{Jeroen~van~den~Brink}
\affiliation{Leibniz Institute for Solid State and Materials Research, IFW Dresden, D-01069, Germany}
\affiliation{Department of Physics, TU Dresden, D-01062 Dresden, Germany}

\author{J.~Wosnitza}
\affiliation{Dresden High Magnetic Field Laboratory (HLD), Helmholtz-Zentrum Dresden-Rossendorf, D-01328, Germany}
\affiliation{Department of Physics, TU Dresden, D-01062 Dresden, Germany}

\author{S.~A.~Zvyagin}
\affiliation{Dresden High Magnetic Field Laboratory (HLD), Helmholtz-Zentrum Dresden-Rossendorf, D-01328, Germany}

\author{I.~Rousochatzakis}
\affiliation{Leibniz Institute for Solid State and Materials Research, IFW Dresden, D-01069, Germany}

\date{\today}

\begin{abstract}
The recent discovery of skyrmions in Cu$_2$OSeO$_3$ has established a new platform to create and manipulate skyrmionic spin textures. We use high-field electron spin resonance (ESR) spectroscopy combining a terahertz free electron laser and pulsed magnetic fields up to 64 T to probe and quantify its microscopic spin-spin interactions. Besides providing direct access to the long-wavelength Goldstone mode, this technique probes also the high-energy part of the excitation spectrum which is inaccessible by standard low-frequency ESR. Fitting the behavior of the observed modes in magnetic field to a theoretical framework establishes experimentally that the fundamental magnetic building blocks of this skyrmionic magnet are rigid, highly entangled and weakly coupled tetrahedra.
\end{abstract}


\maketitle


In recent years there has been an enormous experimental activity in non-centrosymmetric helimagnets~\cite{Muhlbauer2009,tonomura2012,Yu2010,Muenzer2010,yu2012,Seki2012Sc,Adams2012,SekiSANS2012}, where the chiral Dzyaloshinsky-Moriya (DM) interactions~\cite{Dzyaloshinsky,*Moriya} stabilize skyrmions, topological particle-like magnetization textures, originally introduced by Skyrme in the context of subatomic particle physics~\cite{skyrme1962}. As first predicted by Bogdanov and Yablonskii~\cite{Bogdanov1989a,*Bogdanov1989b}, skyrmions may condense spontaneously into a lattice at thermodynamic equilibrium~\cite{Roessler2006}, in analogy to Abrikosov vortices in type-II syperconductors~\cite{Abrikosov1957}, or the ``blue phases'' in cholesteric liquid crystals~\cite{Wright89}. While most of the well-known skyrmionic helimagnets, such as MnSi~\cite{Muhlbauer2009,tonomura2012}, Fe$_{1-x}$Co$_x$Si~\cite{Yu2010,Muenzer2010}, and FeGe~\cite{yu2012} are metallic, the recent discovery~\cite{Seki2012Sc,Adams2012,SekiSANS2012} of skyrmionic mesophases in \cuse, a strongly correlated insulator with localized Cu$^{2+}$ spins~\cite{Kohn1977,Bos2008,Belesi2010}, has opened a route to explore skyrmion physics in Mott insulators. In addition \cuse\ manifests a magnetoelectric coupling~\cite{Seki2012Sc,Belesi2012,Maisuradze2012} which brings exciting perspectives on the application front, since it allows to manipulate skyrmions by an external electric field~\cite{White2012,Lin2013,Fert2013}.

\begin{figure}[!b]
\includegraphics[width=0.47\textwidth]{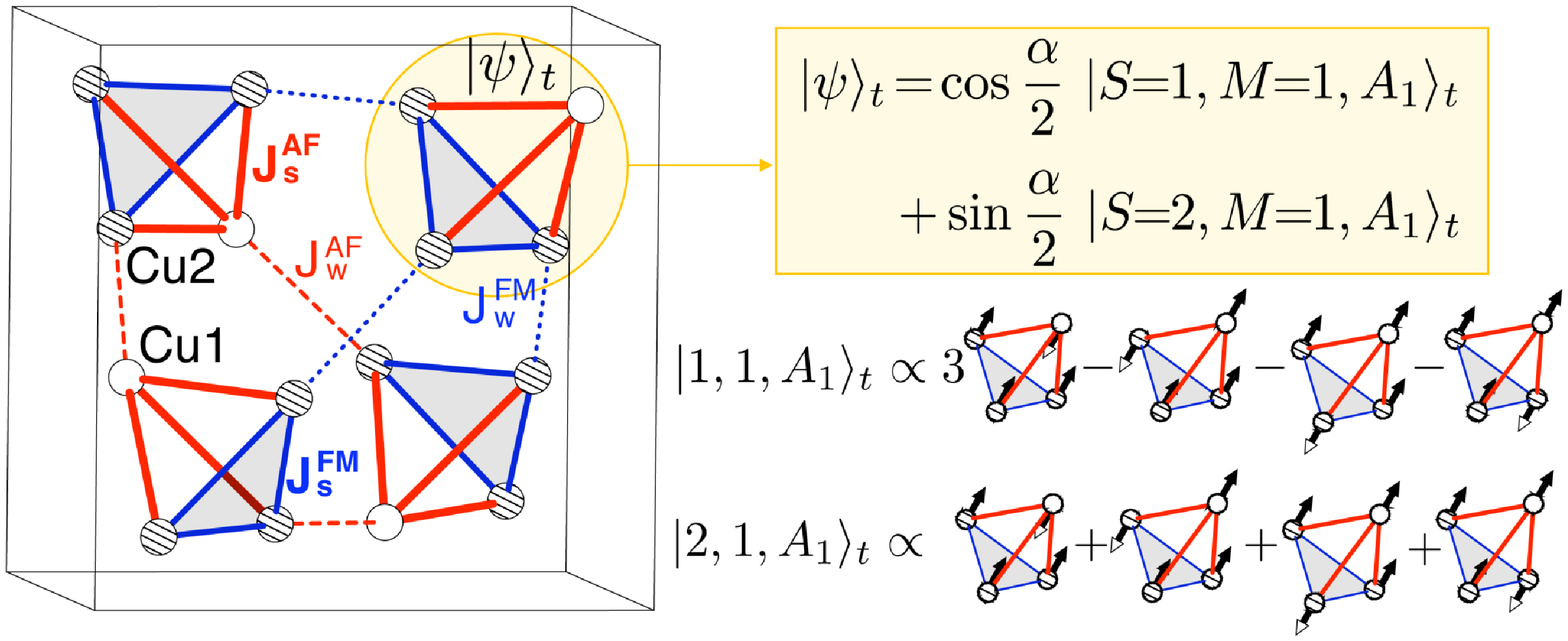}
\caption{\label{fig:1} (Color online)
Distorted pyrochlore structure of the Cu$^{2+}$ ions in \cuse, showing the proposed alternation of magnetically `weak' and `strong' tetrahedra. In total, there are five exchange couplings~\cite{Yang2012,Oleg2014}: two strong, $J^{\sf AF}_{\sf s}$ and $J^{\sf FM}_{\sf s}$ (solid lines), two weak, connecting strong tetrahedra, $J^{\sf AF}_{\sf w}$ and $J^{\sf FM}_{\sf w}$ (dashed lines), and a weak longer-range exchange $J^{\sf AF}_{\sf O..O}$ (not shown). 
The shaded circle indicates a `strong' tetrahedron with wave function  $|\psi\rangle_t$, which strongly deviates from a semiclassical ``3up-1down''  state.
}
\end{figure}

Another very attractive aspect rooted in the insulating nature of \cuse\ is that a reliable modeling of its magnetic interactions becomes possible which, in conjunction with powerful experimental techniques like the one presented below, offers the unique opportunity to gain a precise understanding of the microscopic magnetic structures and interactions in this skyrmionic material. Having a non-centrosymmetric space group $P2_13$, the magnetic Cu$^{2+}$ ions in \cuse\ reside at the vertices of a distorted pyrochlore lattice, featuring two symmetry-inequivalent Cu sites, Cu1 and Cu2, with ratio 1:3 (in total there are 16 Cu sites per unit cell), see Fig.~\ref{fig:1}. While the basic local magnetism is believed to be roughly pictured in terms of a semiclassical 3up-1down structure~\cite{Kohn1977,Bos2008,Belesi2010}, density functional based band-structure calculations suggest that the basic building blocks of helimagnetism are not individual Cu spins but rather quantum-mechanical (QM) tetrahedral spin entities (shaded circle in Fig.~\ref{fig:1}) persisting far above the magnetic ordering temperature of $T_C\!\simeq\!60$~K \cite{Oleg2014}. From the calculations two well-separated exchange energy scales can be distinguished, defined by inter- and intra-tetrahedra exchange coupling interactions (hereafter `weak' and `strong'  interactions, $|J_w|\!\sim\!20$-$50$~K and $|J_s|\!\sim\!100$-$150$~K, respectively, Fig.~\ref{fig:1}). The rigid Cu$_4$ entities are then engendered by the strong couplings $J_s$: the magnetic ground state (GS) of an isolated `strong' tetrahedron is separated from its excited states by a large gap of $\Delta\!\simeq\!280$~K. Such a separation of energy scales identified by the calculations will very strongly affect the long-wavelength physics, the stability and interplay of skyrmionic and half-skyrmionic phases and the magnetoelectricity~\cite{Oleg2014,Romhanyi2014}. Here, we set out to test its presence experimentally.

\begin{figure*}[!t]
\includegraphics[width=0.95\textwidth]{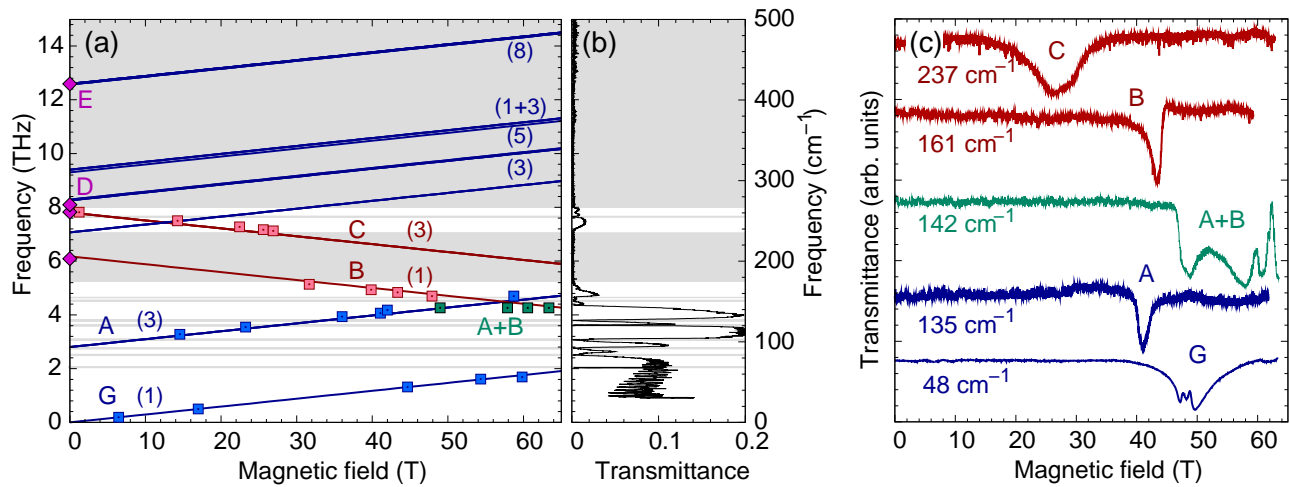}
\caption{\label{fig:2} (Color online)
(a) ESR spectral data for field along [100] and $T\!\simeq\!{6}$~K. Lines are calculations (numbers give the degeneracy of each mode), see Fig.~\ref{fig:3}(c). Diamonds denote the absorption data reported by Raman~\cite{Gnezdilov2010} and far-infrared~\cite{Miller2010} studies, see text. Shaded areas show the regions with very low signal intensity due to strong phonon absorptions, as shown by the zero-field transmittance data at 4.2 K (b). (c) Examples of ESR spectra (the spectra are offset for clarity).
}
\end{figure*}

A direct fingerprint of the presence of weak and strong magnetic interactions will be in the magnetic excitation spectrum. In particular, weak interactions between tetrahedra, which do not alter the integrity of the strongly coupled tetrahedral entities, will cause a Goldstone mode plus a three-fold mode with energy $E\!\propto\!J_w$~\cite{Romhanyi2014}. The strong interactions in turn will cause the presence of a large number (56) of single-particle modes which correspond to local, intra-tetrahedra excitations that appear at higher energies ($E\!\propto\!J_s$). Probing this magnetic mode structure requires not only access to a wide frequency range, reaching up to the far infrared, but also the ability for fine tuning through a forest of strong phonon absorption lines that are present in the same region~\cite{Gnezdilov2010,Miller2010}. Here, we achieve this goal by employing a high-field electron spin resonance (ESR) spectroscopy using pulsed magnetic fields up to 64 T and a terahertz (THz) free electron laser as a high-power monochromatic source in the range 1.2 - 70 THz. As we show below, this powerful technique gives direct access not only to the long-wavelength Goldstone mode, which has also been accessed by low-frequency (GHz-range) ESR~\cite{Larranaga2009, Kobets2010, Maisuradze2012}, but also to the high-energy part of the excitation spectrum. We will show that fitting these ESR lines to the theoretical framework not only provides direct experimental evidence of the tetrahedral picture of \cuse, but also unambiguously quantifies the microscopic exchange couplings. This is especially important for the set of strong coupling parameters $J_s$ which cannot be readily extracted from thermodynamic and magnetization measurements due to the large gap $\Delta$~\footnote{For example, the value of $T_C$ is almost entirely fixed by the weak couplings $J_w$ alone~\cite{Oleg2014}.}.

A high-quality single crystal of Cu$_2$OSeO$_3$ was grown by chemical-vapor transport method. Experiments were performed on a $2\!\times\!3\!\times\!1.42$~mm$^3$ piece using a pulsed-field ESR spectrometer and a THz-range free electron laser at the Helmholtz-Zentrum Dresden-Rossendorf~\cite{spectrometer,Zvyagin2009}. The latter employs a superconducting linear accelerator with pulse repetition rate of 13 MHz. This provides an operation of the THz laser in a quasi-cw mode with the radiation bandwidth of $\sim$0.3-0.7$\%$ and intensity fluctuation less than 5$\%$. This unique facility allows for ESR studies in the wavelength range of 4-250 $\mu$m (which corresponds to 75-1.2 THz or wavenumbers between 2500 and 40 cm$^{-1}$) and in pulsed magnetic fields up to 64 T and beyond~\cite{Zvyagin2009}. VDI diodes (product of Virginia Diodes Inc.) were employed as radiation sources at frequencies below 0.5 THz. The base temperature of the sample was  $\sim 6$~K, that is well below the ordering temperature $T_C$. The magnetic field was applied along the cubic [001] crystallographic axis. We remark here that the magnitude of the applied field is always above the critical field $H_{c2}\!\sim$0.1~T where the transverse, helical component triggered by the DM anisotropy vanishes~\cite{Seki2012Sc,Belesi2012,Adams2012}.

The main experimental findings are shown on the frequency-magnetic field diagram in Fig.~\ref{fig:2}(a). The shading denotes the regions that suffer from low signal intensity due to strong phonon absorptions~(Fig.~\ref{fig:2}(b)). Four ESR modes, G, A, B, and C, were detected in the frequency range between 44 and 270 cm$^{-1}$. In all cases, the relative ESR absorption was up to 50 $\%$ compared to that in zero field. All the modes can be described using the effective $g_{eff}\sim2.1\pm0.1$, typical for Cu-ions. Extrapolating the frequency-field dependence of mode B to zero frequency gives the field $H\!\simeq\!215$~T, where the minority, Cu1 spins are flipped. Extrapolating the frequency-field dependences of modes A, B, and C to zero field, on the other hand, gives the zero-field gaps of 95, 202, and 263 cm$^{-1}$, respectively. The mode G corresponds to the expected Goldstone mode that has also been seen in low-frequency studies~\cite{Larranaga2009, Kobets2010, Maisuradze2012,footnoteG}. 

Now, we would like to notice a fine structure of the resonance absorption observed at 142 cm$^{-1}$ in magnetic fields $\sim\!50$-$60$~T (Fig.~\ref{fig:2}(c)). This fine  structure corresponds to the level crossing of modes A and B, and is consistent with the theoretical prediction that mode A is three-fold degenerate (see below). Another important observation is that mode C (at a frequency of 237 cm$^{-1}$) is much broader than other modes, having a shoulder. The peculiar lineshape of mode C is again consistent with the theory, which predicts a 3-fold degeneracy for this mode. 

\begin{figure}[!t]
\includegraphics[width=0.49\textwidth]{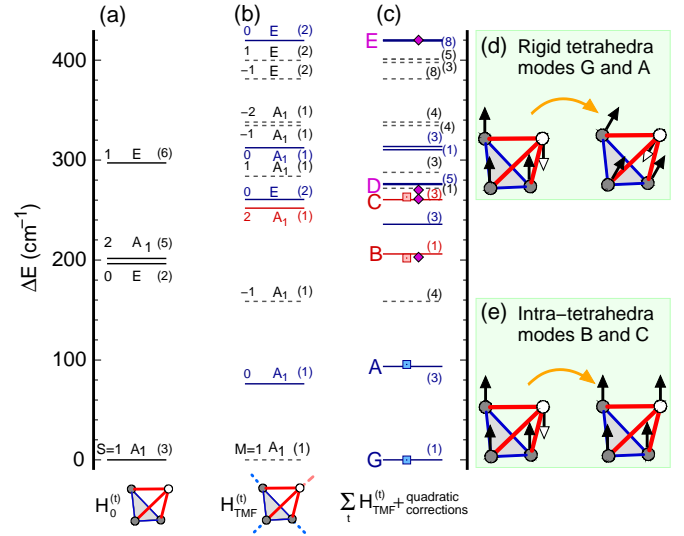}
\caption{\label{fig:3} (Color online)
(a) Spectrum of an isolated `strong' tetrahedron $t$, $\mc{H}_0^{(t)}$, labeled by the IRs of ${\sf C}_{3v}$, the total spin $S$ of $t$ and its projection $M$ along the $\vec{z}$ axis.
(b) Spectrum of $\mc{H}_{\sf TMF}^{(t)}\!=\!\mc{H}_0^{t}\!+\!\mc{V}_{\sf MF}^{(t)}$, where $\mc{V}_{\sf MF}^{(t)}$ includes the exchange fields exerted by neighboring tetrahedra at a mean-field level. Here,  $S$ is not a good quantum number.
(c) Zero-momentum excitations of the full Hamiltonian from quadratic multi-boson theory. Solid lines denote states with $M\!=\!0$ or $2$, that are accessible by ESR.
(d-e) Semiclassical pictures for two representative modes of different type. 
}
\end{figure}

It is interesting to compare the ESR observations to the results from Raman scattering. Gnezdillov {\it et al.} have reported a large number of modes; some of them can be regarded as magnetic excitations ~\cite{Gnezdilov2010}. These modes are shown in Figs.~\ref{fig:2}(a) and \ref{fig:3}(c) by diamonds. The gaps of modes B and C (203 cm$^{-1}$ and 261 cm$^{-1}$) perfectly coincide with our observations, giving unambiguous evidence for their magnetic origin~\footnote{We note that the mode at 86 cm$^{-1}$, observed by means of Raman scattering~\cite{Gnezdilov2010}, was interpreted as the collective vibration of the edge-sharing CuO$_5$ units~\cite{Miller2010}.}. The modes D (270 cm$^{-1}$) and E (420 cm$^{-1}$) are exchange modes (Figs.~\ref{fig:2}(a) and \ref{fig:3}(c)). It is worth mentioning that lines B and D have also been reported in the far-infrared study of Miller {\it et al.}~\cite{Miller2010}.

The theoretical treatment of the system with rigid tetrahedral entities that are weakly coupled can be achieved by expanding around a reference variational wave-function $|\Psi\rangle\!=\!\prod_t^{\otimes} |\psi\rangle_t$, which is the tensor product over the tetrahedral entities $|\psi\rangle_t$, see Fig.~\ref{fig:1} \cite{Romhanyi2014}. The latter can be found self-consistently by diagonalizing the mean-field Hamiltonian $\mc{H}_{\sf TMF}^{(t)}\!=\!\mc{H}_0^{(t)}\!+\!\mc{V}_{\sf MF}^{(t)}$, where $\mc{H}_0^{(t)}$ contains the strong couplings $J_s$ inside the tetrahedron $t$, and $\mc{V}_{\sf MF}^{(t)}$ contains the self-consistent mean fields exerted by neighboring tetrahedra $t'$ on $t$, via the weak couplings $J_w$; 
The explicit form of these Hamiltonians can be found in the supplement. The remaining interactions, described by $\mc{H}\!-\!\sum_t\mc{H}_{\sf TMF}^{(t)}$, can then be treated using a multi-boson generalization of the standard spin-wave expansion. The main effect of these interactions is to weakly renormalize the energies and give rise to a finite dispersion. The correction to the GS expectation values of the local spin lengths is about 1\% or less~\cite{Romhanyi2014}, showing that the GS correlations are captured almost entirely by the tetrahedra-factorised wavefunction $|\Psi\rangle$ and that the method is internally consistent.

The resulting spectra of $\mc{H}_0^{(t)}$ and $\mc{H}_{\sf TMF}^{(t)}$ for an isolated tetrahedron, as well as the final multi-boson spectrum of the full Hamiltonian in the zero-momentum sector are shown, respectively, in Figs.~\ref{fig:3}(a), (b), and (c).  Here the exchange couplings have been fitted to the experimental data (see details below). From the symmetries of the calculated wave functions we can extract in detail the nature of the different magnetic modes. The spectrum of $\mc{H}_0^{(t)}$ is classified in terms of the total spin $S$ of the tetrahedron, its projection $M$ along the $\vec{z}$ axis, as well as the irreducible representations (IRs) of the point group ${\mathsf C}_{3v}$. The GS is a symmetric (${\sf A}_1$) triplet, followed by a pair of singlets belonging to the two-dimensional sector ${\sf E}$, an ${\sf A}_1$ quintet, and finally a pair of triplets belonging to ${\sf E}$ (see supplement). Turning on the weak couplings $J_w$ at the mean-field level does not lower the point group and the total $M$ remains a good quantum number, but this is not the case for the total spin $S$. As a result, the Zeeman degeneracy is lifted and, in addition, we have a direct coupling between levels belonging to the same $M$ and the same IR of ${\sf C}_{3v}$. Most notably, the three components of the GS triplet $\mc{H}_0^{(t)}$ split and, in addition, they admix a finite portion of the corresponding components of the ${\sf A}_1$ quintet, see formula for $|\psi\rangle_t$ in Fig.~\ref{fig:1}. The remaining levels are affected analogously, see Fig.~\ref{fig:3}(b).

We are now ready to analyze the full excitation spectrum of Fig.~\ref{fig:3}(c), building on the main features of the spectrum of $\mc{H}_{\sf TMF}^{(t)}$ and by keeping in mind that there are four tetrahedra per unit cell. For concreteness, we shall only focus on the four lowest modes G, A, B, and C that are observed in our ESR experiments. These modes are dipolar, i.e., they correspond to transitions from $|\psi\rangle_t$, where $M\!=\!1$, to states with $M'\!=\!0$ or $2$. Labeling the eigenstates of $\mc{H}_{\sf TMF}^{(t)}$ by an index $n\!=\!1$-$16$, with $n\!=\!1$ corresponding to $|\psi\rangle_t$, there is one excitation per tetrahedron from $n\!=\!1$ to $n'\!=\!2$, which in total gives four excitations with $\delta M\!=\!M'\!-\!M\!=\!-1$. The Goldstone mode G corresponds to the uniform combination, while the remaining three modes give the mode A, which belongs to the 3-dimensional sector ${\sf T}$ of the space group $P2_13$ of the wavevector $\vec{k}\!=\!0$. Physically, the finite energy of this mode amounts to the energy cost of rotating one of the four tetrahedral entities against the mean field exerted by its neighbors. Importantly, the physical content of the rotated tetrahedron is not altered since the state $|n'\!=\!2\rangle\!\propto\! S^-|\psi_t\rangle$, i.e., it is simply a global rotation of $|\psi\rangle_t$.

The origin of modes B and C is fundamentally different, as they correspond to the four possible ways to excite from $n\!=\!1$ to $n'\!=\!6$, which is a state that originates from the fully polarized (quintet) state of $\mc{H}_0^{(t)}$. In other words, this mode corresponds (semi-classically) to rotating the minority spin inside a single tetrahedron, which costs an energy $E\!=\!2J_s^{\sf AF}$ at the level of $\mc{H}_0^{(t)}$. 
Again, there are four such modes per unit cell which, under the full space group of the $\Gamma$ point, split into 
a 1-fold ({\sf A}$_1$) and a 3-fold ({\sf T}) mode, corresponding, respectively, to modes B and C. In contrast to modes G and A which have $\delta M\!=\!-1$ and thus become harder in a magnetic field, B and C have $\delta M\!=\!+1$ and thus become softer in a field. Incidentally, B is the first mode that condenses at larger fields, marking the transition to the fully polarized state, as mentioned above. We also note in passing that the modes A and C do not belong to the symmetric representation ${\sf A}_1$ of $P2_13$, and so their detection by ESR is possible only due to anisotropy and the symmetry inequivalence of the electronic $\vec{g}$-tensors of the two Cu sites.

Besides clarifying the nature of the observed modes, their multiplicity and their behavior under a magnetic field, we also obtain direct information on the exchange couplings. As a guiding principle, we keep the weak-coupling values of \cite{Oleg2014} intact (namely, $J^{{\sf AF}}_w\!=\!27$~K, $J^{{\sf FM}}_w\!=\!-50$~K, and $J^{{\sf AF}}_{{\sf O..O}}\!=\!45$~K). Since these couplings directly determine the system's thermodynamic properties they can be considered fixed by fits to experimental magnetization data~\cite{Oleg2014}. In fact, the striking agreement in the energy of our ESR mode A, which depends practically only on $J_w$, gives a direct experimental verification of these values. The strong coupling $J_s^{\sf AF}$ can be extracted from the energies of modes B and C, which depend only on this parameter (see supplementing material), giving us $J^{{\sf AF}}_s\!\simeq\!145$~K. The second strong coupling, $J_s^{\sf FM}$,  can be extracted by fitting the mode E (observed by Raman scattering~\cite{Gnezdilov2010}) to the highest excitation originating from the two-fold triplet state of Fig.~\ref{fig:3}(a), which fixes $J_s^{\sf FM}\!\simeq\!-140$~K. These values are comparable to the calculated ones of 170~K and -128~K \cite{Oleg2014}, be it 10-15\% lower.

In summary, we have performed a high-field resonance study of the chiral helimagnet \cuse\ using pulsed magnetic fields up to 64 T and a terahertz-range free electron laser as a high-power monochromatic source, enabling fine tuning over a broadband frequency range and through a forest of strong phonon absorption regions. Apart from detecting the long-wavelength Goldstone excitation, several higher energy modes are disclosed. Comparing these results to a theoretical framework clarifies the detailed nature of the observed modes and allows to unambiguously extract the magnetic interaction parameters. These findings establish experimentally that the fundamental magnetic building blocks of \cuse\ are rigid, highly entangled tetrahedra, on account of a large separation of exchange energy scales. The latter  has been shown to dramatically affect the long-wavelength helimagnetism, the presence of antiferromagnetic canting modes, the interplay of skyrmionic and half-skyrmionic phases~\cite{Oleg2014}, as well as the magnetoelectricity~\cite{Romhanyi2014}.

{\it Acknowledgements} ---
We thank Wolfgang Seidel for his help during operation of the free-electron laser. We acknowledge the support of the HLD-HZDR, member of the European Magnetic Field Laboratory (EMFL). S.A.Z. appreciates the support of the DFG foundation. 

\bibliographystyle{apsrev4-1}
\bibliography{Cu2OSeO3_ESR}

\onecolumngrid


\section {Supplementary Material}
\twocolumngrid
\subsection{Eigensystems of $\mathcal{H}_0$ and $\mathcal{H}_{\sf TMF}$}
Here, we provide explicit details for the eigenspectrum of the single-tetrahedron Hamiltonians $\mc{H}_0^{(t)}$ and $\mc{H}_{\sf TMF}^{(t)}$. Denoting by $\vec{S}_1$ and $\vec{S}_{2-4}$ the Cu1 and the three Cu2 spins of the tetrahedron $t$, respectively, we have:
\bea
\mc{H}_0^{(t)}&=&
J_s^{\sf AF} \vec{S}_1\cdot \vec{S}_{234}
+J_s^{\sf FM} \left( \vec{S}_2\cdot\vec{S}_3 + \vec{S}_3\cdot\vec{S}_4+\vec{S}_4\cdot\vec{S}_2 \right)
\nonumber\\
&=&
\frac{J_s^{\sf AF}}{2} \vec{S}^2
+\frac{J_s^{\sf FM}-J_s^{\sf AF}}{2}\vec{S}_{234}^2 + c~,
\eea
where the constant $c\!=\!-3\left(J_s^{\sf AF}\!+\!3J_s^{\sf FM}\right)/8$, and we have defined $\vec{S}_{234}\!=\!\vec{S}_2\!+\!\vec{S}_3\!+\!\vec{S}_4$, and the total spin $\vec{S}\!=\!\vec{S}_1\!+\!\vec{S}_{234}$. So, the eigenspectrum can be found analytically by standard addition of angular momenta, in terms of the spin quantum numbers $S_{23}$, $S_{234}$, $S$, and the projection $M$ along the quantization $\vec{z}$ axis. In addition, the eigenstates can be labeled by irreducible representations $\lambda$ of the point group C$_{3v}$ of $\mc{H}_0^{(t)}$. The results are given in Table~\ref{tab:H0}, where each eigenstate is labeled as $|S_{23},S_{234},S,M\rangle_\lambda$. For convenience, all energies are measured  from the energy $E_0\!=\!\frac{-5J_s^{\sf AF}+3J_s^{\sf FM}}{4}$, of the ground state triplet $|1,3/2,1,M\rangle_{A_1}$.

Turning to the tetrahedral mean-field Hamiltonian, this is given by 
\bea
\mc{H}_{\sf TMF}^{(t)}&=&\mc{H}_0^{(t)}
+J_s^{\sf AF}\left(\delta~S^z_1+\Delta~S^z_{234}\right)\;,
\eea
where the mean-field parameters $\delta$ and $\Delta$ stand for the Weiss fields (in units of $J_s^{\sf AF}$) exerted from the neighboring tetrahedra and thus depend in a self-consistent way on the expectation values of the magnetizations $\langle S^z_1\rangle$ and $\langle S^z_2\rangle$ along the quantization axis. Counting the number of inter-tetrahedra couplings of each type gives
\bea
\delta &=& 3\langle S^z_2\rangle(J^{\sf AF}_w+J^{{\sf AF}}_{{\sf O..O}})/J_s^{\sf AF} > 0~, \nonumber\\
\Delta &=& \left[\langle S^z_1\rangle(J^{\sf AF}_w+J^{{\sf AF}}_{{\sf O..O}})+2\langle S^z_2\rangle J^{\sf FM}_w\right]/J_s^{\sf AF} < 0~. 
\eea
As discussed in the main text, these extra terms do not lower the point-group symmetry but they admix states with different $S$. In particular, the ground state of $\mc{H}_{\sf TMF}^{(t)}$ is a combination of the triplet ground state and the quintet state of $\mc{H}_0^{(t)}$, see Fig.~1 of main text:
\bea
|\Psi\rangle_{\sf TMF}=\cos\frac{\alpha}{2}|1,3/2,1,1\rangle_{{\sf A}_1}+\sin\frac{\alpha}{2}|1,3/2,2,1\rangle_{{\sf A}_1}~.~~~
\eea
The mixing parameter $\alpha$ controls the magnetization on the Cu1 and Cu2 sites, 
\be
\langle S^z_1\rangle=-\frac{\cos\alpha-\sqrt{3}\sin\alpha}{4},~~
\langle S^z_2\rangle=\frac{1-\langle S^z_1\rangle}{3}~, 
\ee
and depends on the Weiss fields in a self-consistent way:
\bea
\tan\frac{\alpha}{2}&\!=\!&
\frac{\sqrt{3}(\delta-\Delta)}{4\!+\!\delta\!-\!\Delta
\!+\!4\sqrt{1\!+\!\frac{\delta\!-\!\Delta}{2}+(\frac{\delta\!-\!\Delta}{2})^2
}}~.
\eea
For the couplings $J_s^{\sf AF}\!=\!145$~K, $J_s^{\sf FM}\!=\!-140$~K, $J_w^{\sf AF}\!=\!27$~K, $J_w^{\sf FM}\!=\!-50$~K, and $J^{\sf AF}_{\sf O..O}\!=\!45$~K, the optimal variational parameter is $\alpha= 0.384856$.

Table~\ref{tab:HTMF} shows the energies of the 16 eigenstates of $\mc{H}_{\sf TMF}^{(t)}$, measured from the ground-state energy $E_0$ of $\mc{H}_0^{(t)}$. We also give the energies up to the leading, linear order in $\delta$ and $\Delta$.

\begin{table}[!t]
\caption{The eigenspectrum of $\mc{H}_0^{(t)}$, classified in terms of the spin quantum numbers $S_{23}$, $S_{234}$, $S$, the projection $M$ along the quantization axis as well as the symmetry sectors $\lambda$ of the point group C$_{3v}$, see text.\label{tab:H0}}
\begin{ruledtabular}
\begin{tabular}{cccc}
\raisebox{0.3ex}[0pt]{n}&
\raisebox{0.3ex}[0pt]{$|S_{23},S_{234},S,M\rangle_\lambda$} &
\raisebox{0.3ex}[0pt]{M}&
\raisebox{0.3ex}[0pt]{$E-E_0$}\\ 
\hline
\noalign{\smallskip}
$1$ & 
$|1,3/2,1,1\rangle_{{\sf A}_1} $& $1$ & {} \\
$2$  & 
$|1,3/2,1,0\rangle_{{\sf A}_1} $ & ${\color{darkblue} 0}$ & 0\\ 
$3$ &
$|1,3/2,1,\overline{1}\rangle_{{\sf A}_1} $ & $-1$ & {}\\ 
\noalign{\smallskip}
\hline
\noalign{\smallskip}
$\begin{array}{c} 4 \\ 5\end{array}$ & 
$\left.\begin{array}{c} |1,1/2,0,0\rangle_{{\sf E}_1} \\ |0,1/2,0,0\rangle_{{\sf E}_2} \end{array}\right\}$ & ${\color{darkblue} 0}$ & 
$\frac{1}{2}(J^{\sf AF}_s-3J^{\sf FM}_s)$\\
\noalign{\smallskip}
\hline
\noalign{\smallskip}
$6$ &
$|1,3/2,2,\overline{2}\rangle_{{\sf A}_1} $ & $-2$ & {} \\ 
$7$  &
$|1,3/2,2,\overline{1}\rangle_{{\sf A}_1} $ & $-1$ & {}\\  
$8$ & 
$|1,3/2,2,0\rangle_{{\sf A}_1} $ & ${\color{darkblue}0}$ & $
2J^{\sf AF}_s$\\
$9$ &
$|1,3/2,2,1\rangle_{{\sf A}_1} $ & $1$ & {}\\
$10$ &
$|1,3/2,2,2\rangle_{{\sf A}_1} $ & ${\color{darkred}2}$ & {}\\
\noalign{\smallskip}
\hline
\noalign{\smallskip}
$\begin{array}{c} 11 \\ 12 \end{array}$ &
$\left.\begin{array}{c} |1,1/2,1,\overline{1}\rangle_{{\sf E}_1} \\ |0,1/2,1,\overline{1}\rangle_{{\sf E}_2} \end{array}\right\}$ & $-1$ & {}\\
\noalign{\smallskip}
$\begin{array}{c} 13 \\ 14 \end{array}$ & 
$\left.\begin{array}{c} |1,1/2,1,0\rangle_{{\sf E}_1} \\ |0,1/2,1,0\rangle_{{\sf E}_2} \end{array}\right\}$ & ${\color{darkblue}0}$ & $\frac{3}{2}(J^{\sf AF}_s-J^{\sf FM}_s)$\\
\noalign{\smallskip}
$\begin{array}{c} 15 \\ 16 \end{array}$ &
$\left.\begin{array}{c} |1,1/2,1,1\rangle_{{\sf E}_1} \\ |0,1/2,1,1\rangle_{{\sf E}_2} \end{array}\right\}$ & $1$ & {}\\
\noalign{\smallskip}
\end{tabular}
\end{ruledtabular}
\end{table}

\begin{table}[!t]
\caption{
The eigenspectrum of $\mc{H}_{\sf TMF}^{(t)}$, classified in terms of the total spin projection $M$ along the quantization axis and the symmetry sectors $\lambda$ of the point group C$_{3v}$, see text. The symbol $j_f$ stands for the ratio $j_f=J_s^{\sf FM}/J_s^{\sf AF}$.
\label{tab:HTMF}}
\begin{ruledtabular}
\begin{tabular}{cccc}
\raisebox{0.3ex}[0pt]{$\tilde n$}&
\raisebox{0.3ex}[0pt]{$\lambda$} &
\raisebox{0.3ex}[0pt]{M}&
\raisebox{0.3ex}[0pt]{$(E_{\sf TMF}-E_0)/J_s^{\sf AF}$}\\ 
\hline
\noalign{\smallskip}
\noalign{\smallskip}
$1$ & ${\sf A}_1 $& $1$ & 
$1 +\Delta-[1+\frac{\delta-\Delta}{2}+(\frac{\delta-\Delta}{2})^2]^{1/2}\simeq \frac{-\delta+5\Delta}{4}$ \\
\noalign{\smallskip}
$2$  & ${\sf A}_1 $ & ${\color{darkblue} 0}$ &
$1 -[1+(\frac{\delta-\Delta}{2})^2]^{1/2}\simeq 0$\\
\noalign{\smallskip}
$3$ &${\sf A}_1 $ & $-1$ & 
$1 -\Delta-[1+\frac{\delta-\Delta}{2}+(\frac{\delta-\Delta}{2})^2]^{1/2}\simeq-\frac{\delta+3\Delta}{4}$\\
\noalign{\smallskip}
\noalign{\smallskip}
\hline
\noalign{\smallskip}
\noalign{\smallskip}
$\begin{array}{l} 4 \\ 5\end{array}$ & ${\sf E}$ & ${\color{darkblue} 0}$ & 
$1-\frac{3}{2}j_f-\frac{1}{2}[1+(\delta-\Delta)^2]^{1/2}\simeq \frac{1-3j_f}{2}$\\
\noalign{\smallskip}
\hline
\noalign{\smallskip}
\noalign{\smallskip}
$6$ &${\sf A}_1$ & $-2$ & 
$2-\frac{\delta+3\Delta}{2}$ \\
\noalign{\smallskip}
$7$  &${\sf A}_1 $ & $-1$ & 
$1-\Delta+[1+\frac{\delta-\Delta}{2}+(\frac{\delta-\Delta}{2})^2]^{1/2}\simeq 2+\frac{\delta-5\Delta}{4}$\\
\noalign{\smallskip}
$8$ & ${\sf A}_1 $ & ${\color{darkblue}0}$ & 
$1+[1+(\frac{\delta-\Delta}{2})^2]^{1/2}\simeq 2$\\
\noalign{\smallskip}
$9$ & ${\sf A}_1 $ & $1$ & 
$1+\Delta+[1+\frac{\delta-\Delta}{2}+(\frac{\delta-\Delta}{2})^2]^{1/2}\simeq2+\frac{\delta+3\Delta}{4}$\\
\noalign{\smallskip}
$10$ & ${\sf A}_1 $ & ${\color{darkred}2}$ & 
$2+\frac{\delta+3\Delta}{2}$\\
\noalign{\smallskip}
\noalign{\smallskip}
\hline
\noalign{\smallskip}
\noalign{\smallskip}
$\begin{array}{c} 11 \\ 12 \end{array}$ & ${\sf E}$ & $-1$ & 
$\frac{3}{2}(1-j_f)-\frac{\delta+\Delta}{2}$\\
\noalign{\smallskip}
\noalign{\smallskip}
$\begin{array}{c} 13 \\ 14 \end{array}$ & ${\sf E}$ & ${\color{darkblue}0}$ & 
$1-\frac{3}{2}j_f+\frac{\sqrt{1+(\delta-\Delta)^2}}{2}\simeq\frac{3}{2}(1-j_f)$\\
\noalign{\smallskip}
\noalign{\smallskip}
$\begin{array}{c} 15 \\ 16 \end{array}$ & ${\sf E}$ & $1$ & 
$\frac{3}{2}(1-j_f)+\frac{\delta+\Delta}{2}$\\
\noalign{\smallskip}
\end{tabular}
\end{ruledtabular}
\end{table}

\subsection{Influence of strong couplings on the spectrum}
As explained in detail in \cite{Oleg2014}, the strong couplings $J_s^{\sf AF}$ and $J_s^{\sf FM}$ do not enter the low-energy effective description of Cu$_2$OSeO$_3$. In part, this is reflected in the fact that the value of $T_C$ depends only on the weak couplings $J_w$. For the same reason, the strong couplings do not affect the energy of the three-fold mode A. However, they are expected to affect the higher-energy, intra-tetrahedra modes. Here, we discuss the explicit form of this dependence and uncover the optimal values of the strong couplings that fit the experimental data.

To this end, we can obtain important insights from the explicit expressions of Table~\ref{tab:HTMF} for the eigenergies of $\mc{H}_{\sf TF}^{(t)}$. These energies  give a very good approximation to the strong coupling dependence of the final excitations energies, since the remaining corrections from the multi-boson fluctuations are basically driven by the weak, inter-tetrahedra couplings. We shall come back and demonstrate this point numerically below.

The mode A corresponds to the transition from $\tilde{n}=1$ to $\tilde{n}'=2$, and the corresponding energy difference does not depend on the strong couplings.
\begin{figure}[t!]
\includegraphics[width=0.48\textwidth]{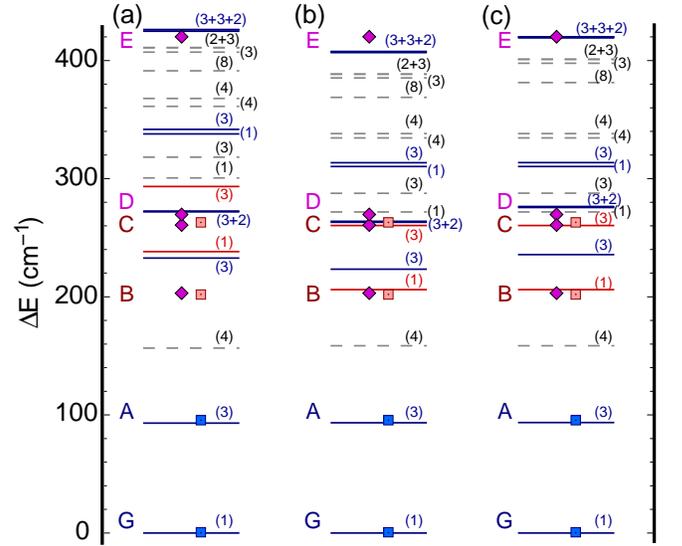}
\caption{\label{fig:spectra} (Color online) Zero-momentum excitations of the quadratic multi-boson Hamiltonian (see main text), using three different sets of strong couplings. 
(a) Strong couplings taken from \cite{Oleg2014}:  $J_s^{\sf AF}=170$ K and $J_s^{\sf FM}=-128$ K. 
(b) To fit the ESR mode B, we adjust the strong antiferromagnetic coupling to $J_s^{\sf AF}=145$ K. 
(c) To fit the Raman mode E, we finally adjust the strong ferromagnetic coupling to $J_s^{\sf FM}=-140$ K. 
Experimental ESR (present work) and Raman~\cite{Gnezdilov2010} data are denoted by squares and diamonds, respectively.}
\end{figure}

%
Turning to modes B and C, these correspond to transitions from $\tilde{n}=1$ to $\tilde{n}'=10$, and the energy difference depends only on $J_s^{\sf AF}$ and not on $J_s^{\sf FM}$. 
Finally, the highest energy dipolar mode corresponds to the transition from $\tilde{n}=1$ to $\tilde{n}'=13-14$, and the energy difference now depends on both $J_s^{\sf AF}$ and $J_s^{\sf FM}$.

To check the effect of multi-boson fluctuations and eventually fit the values of the strong couplings we  compare three multi-boson spectra from three different sets of strong coupling constants. In all cases, the weak couplings are fixed to the values reported in \cite{Oleg2014}: $J_w^{\sf AF}\!=\!27$~K, $J_w^{\sf FM}\!=\!-50$~K, and $J^{\sf AF}_{\sf O..O}\!=\!45$~K. 
In the first set, which gives the spectrum shown in Fig.~\ref{fig:spectra} (a), the strong couplings are again the ones obtained by the {\it ab initio} study of \cite{Oleg2014}: $J_s^{\sf AF}=170$ K and $J_s^{\sf FM}=-128$ K. We see that the observed mode $A$ is in nearly perfect agreement with the calculations, but the remaining higher-energy modes are not.

As discussed above and seen explicitly in Table~\ref{tab:H0} and \ref{tab:HTMF},  the energy of modes B and C do not depend on $J_s^{\sf FM}$, which suggests that we may adjust $J_s^{\sf AF}$ to fit these modes. Indeed, rescaling $J_s^{\sf AF}$ to $145$~K (but keeping $J_s^{\sf FM}=-128$~K) gives the spectrum of Fig.~\ref{fig:spectra}(b), which now gives a nearly perfect agreement with the modes B and C, without spoiling the fit to mode A.

To fix the remaining strong coupling, $J_s^{\sf FM}$, we assume that the mode E observed by Raman scattering in \cite{Gnezdilov2010} corresponds to the highest energy dipolar mode of our spectra, discussed above. To fit this mode we finally adjust $J_s^{\sf FM}=-140$~K, resulting in  the spectrum shown in Fig.~\ref{fig:spectra}(c). We have now fitted mode E without spoiling the fit to the lower-energy modes.
This concludes that extracting the strong couplings from the experimental data can be achieved in a systematic and well-controlled way.

\bibliographystyle{apsrev4-1}

\end{document}